\documentclass[12pt,preprint]{aastex}

\newcommand\Mx{\mbox{$M_{\rm x}$}}
\newcommand\qx{\mbox{$q_{\rm x}$}}
\newcommand\phiml{\mbox{$\phi_{\rm ml}$}}
\newcommand\cs{\mbox{$c_{\rm s}$}}
\newcommand\vk{\mbox{$V_{\rm k}$}}
\newcommand\velr{\mbox{$v_{\rm r}$}}

\newcommand\Td{\mbox{$T_{\rm d}$}}
\newcommand\Teff{\mbox{$T_{\rm eff}$}}

\newcommand\Porb{\mbox{$P_{\rm orb}$}}
\newcommand\Tvis{\mbox{$T_{\rm vis}$}}
\newcommand\Tres{\mbox{$T_{\rm res}$}}
\newcommand\Tml{\mbox{$T_{\rm ml}$}}

\newcommand\rtrunc{\mbox{$r_{\rm trunc}$}}
\newcommand\dlo{\mbox{$d_{\rm L1}$}}
\newcommand\rlo{\mbox{$r_{\rm L1}$}}
\newcommand\Mwdot{\mbox{$\dot{M}_{\rm w}$}}

\newcommand\Mf{\mbox{$M_{\rm f}$}}
\newcommand\Lx{\mbox{$L_{\rm x}$}}
\newcommand\LEdd{\mbox{$L_{\rm Edd}$}}

\shorttitle{Be/black-hole Binaries} \shortauthors{zf et al.}

\begin{document}

\title{Where Are
Be/black-hole Binaries?    }
\author{Fan Zhang, X.-D. Li and Z.-R. Wang}
\affil{Department of Astronomy, Nanjing University, Nanjing
210093, China} \email{zfastro@nju.edu.cn; lixd@nju.edu.cn;
zrwang@nju.edu.cn}

\begin{abstract}
We apply the tidal truncation model proposed by \citet{neg01} to
arbitrary Be/compact star binaries to study the truncation
efficiency dependance on the binary parameters. We find that the
viscous decretion disks around the Be stars could be truncated
very effectively in narrow systems. Combining this with the
population synthesis results of Podsiadlowski, Rappaport and Han
(2003) that binary black holes are most likely to be born in
systems with orbital periods less than about 30 days, we suggest
that most of the Be/black-hole binaries may be transient systems
with very long quiescent states. This could explain the lack of
observed Be/black-hole X-ray binaries. We also discuss the
evolution of the Be/black-hole binaries and their possible
observational features.
\end{abstract}

\keywords{ stars: circumstellar matter, Be - binaries: close -
black hole - X-ray: stars, bursts }

\section{Introduction}

X-ray binaries consist of a neutron star (NS) or a black hole (BH)
accreting from strong stellar winds or via Roche lobe overflow of
the companion star. Depending on the masses of the optical
companions, they are conventionally divided into low- and
high-mass X-ray binaries. By measuring the radial velocity curve
of the non-degenerate donor, the value of the mass function can be
determined, which provides a lower limit on the mass of the
accreting object. When it is combined with the information of the
spectra of the donor and the orbital light curves, actual
measurements of the mass can be obtained. In this way, currently
the masses of 18 compact objects in X-ray binaries have been found
to exceed the the maximum mass possible for a neutron star. These
binaries are thought to be black-hole X-ray binaries (BHXBs) (see
\citet{mcc03} and references therein).

Among the 18 BHXBs, 3 are persistently bright X-ray systems and 15
are X-ray novae in which 6 show recurrent outbursts. All of the
X-ray novae are low-mass X-ray binaries (LMXBs). The three
persistent sources, i.e. Cyg X$-1$, LMC X$-1$ and LMC X$-3$ are
high-mass X-ray binaries (HMXBs) with massive O/B companion stars.
In the most recent catalogue of high mass X-ray binaries (HMXBs)
edited by \citet{liu00}, only 20 out of 130 HMXBs are O/B
supergiant systems including the black hole binaries. Though
roughly two thirds of HMXBs are Be/X-ray binaries, and X-ray
pulsations have been found in about sixty Be/X-ray binary systems
\citep{zio02}, there are no acknowledged Be/BH binaries. The
existence of Be/BH binaries becomes an open problem. The solutions
may be related to two questions: Do they exist? if yes, how can
they be observed?

The first question is connected with the birth-rate estimation of
BH/massive X-ray binaries (BMXBs). Using Monte-Carlo simulation,
\citet{rag99} calculated the number and distribution of binary BHs
with Be stars. They obtained the expected number of Be/BH binaries
to be of order of unity per $20-30$ Be/NS systems. The Be/BH
binaries were found to be highly eccentric with orbital periods
lying in a range of 10 days to several years. The actual number of
Be/BH binaries could be even higher since in their work, the lower
limit of $50 \, M_{\sun}$ for the progenitor of the BH was
adopted, based on the mass of the supergiant Wray 977
\citep{kap95}. But current stellar evolution models suggest that
stars with mass higher than about $20-25\, M_{\sun}$ will result
in BHs\citep{fry99, fry01}. More recently \citet{pod03} did the
binary population synthesis calculation with the mass of the BH
progenitors in the range of $25\, M_{\sun}<M_{\rm p}<45\,
M_{\sun}$, and presented the calculated distribution of the
orbital period and donor mass for BH binaries in their Fig.2. It
is shown that BHs have large possibility to be formed in
relatively narrow systems with orbital periods of less than tens
of days. The formation rate of BMXBs estimated from their result
is about $10^{-5}\ \rm {yr}^{-1}$. Although the massive donor
stars are not constrained to be Be stars in their work, the
conclusion of the orbital distribution may also be applied to
Be/BH binaries, since the fraction of the Be stars to the massive
stars is not small (see \citet{coe00} and references therein).
Nevertheless, theoretical estimation of this fraction is
difficult. It is not yet clear how Be stars eject the
circumstellar envelope. While rapid rotation plays a part in disc
creation, it is not the sole cause \citep{por99}. Whether the
physical properties of the donor star (metallicity, instability in
the photosphere induced by binary rotation, and etc.) have effects
on producing the dense disk winds is still unknown \citep{zam01}.

The observation of supergiant B[e]/BH system, e.g. XTE
J$0421+560$/CI Cam \citep{rob02}, may provide a further proof of
the existence of Be/BH binaries, since theoretically sgB[e]/X-ray
binaries would have less probability to exist than Be X-ray
binaries due to the much larger mass of sgB[e] stars. Thus, the
absence of Be/black-hole binaries might be due to some
observational selection effects. In this work, we try to find the
clue of the second question, that is, why we can hardly observe
them. We would start from the interaction between a Be star and
its compact companion.

Be/X-ray binaries are characterized by their transient nature.
Recently, theoretical work on disc truncation in Be/X-ray binaries
\citep{oka01,oka02} was invoked to explain the X-ray outbursts in
Be/X-ray binaries, and has got considerable success in the
application to several systems \citep{oka01}, especially 4U
0115+63/V635 Cas \citep{neg01} and A0535+26 \citep{hai03}. The key
idea is as follows: The neutron star exerts a net negative tidal
torque on the viscous decretion disk of the Be star, diminishing
the action of the viscous torque outside some critical radius, and
thus resulting in the truncation of the disk. The disc matter
would then accumulate in the outer rings of the disk till
overcoming of the truncation by the effects of global one-armed
oscillations or disk warping, etc. The following sudden infall of
high density disk matter onto the neutron star causes type II
X-ray outbursts. During the episodes between the outbursts, the
neutron star could hardly be observed because of low mass
accretion rate or the propeller effect. On the other hand, if the
tidal truncation is not very efficient, and the disc can extend
beyond the Roche lobe of the Be star at periastron, the matter
could be accreted onto the neutron star during the periastron
passage, resulting in (quasi-)periodic type I bursts.

It is interesting to see what would happen when the neutron star
in the above picture is replaced by a black hole. We extend the
application of this theory to arbitrary systems to find the
influence of mass ratio, orbital period, eccentricity and
viscosity coefficient on the efficiency of disc truncation. The
model and the necessary parameters are briefly described in
section 2. The results are shown in section 3. Our discussion and
conclusions are made in section 4.

\section{Model}

\citet{neg01} suggest a scenario for the behavior of Be/X-ray
binaries based on long-term multiwavelength monitoring and two
theoretical models: (1) the viscous decretion disk of Be stars
proposed by \citet{lee91}, where the mass flow is outward in
contrast with that in accretion disk, and angular momentum of the
disk is added by viscous torque $\Tvis$; (2) the tidal truncation
model investigated by \citet{art94} for the gravitational
interaction of an eccentric binary system with circumstellar or
circumbinary disks.

Following \citet{neg01}, we consider the picture of the Be/X-ray
binary in which the compact star of mass $\Mx$ moves around a Be
star of mass $M_{\ast}$ and radius $R_{\ast}$ in an orbit of
eccentricity $e$, exerting resonant torques $\Tres$ on the Be
decretion disk. As first approximation, the disc is assumed to be
coplanner. The criterion for the disc truncation at a given
resonance radius $\rtrunc$ is
\begin{equation}\label{trunc}
\Tvis+\Tres \leq 0,
\end{equation}
where $\Tres$ can be easily shown to be dominated by the torque of
the inner Lindblad resonance $\Tres\simeq \sum_{ml}(\Tml)_{\rm
ILR}$. For near-Keplerian discs, given the expression of viscosity
and resonance torques as Eqs.(6) and (8) in \citet{neg01}, the
above truncation criterion can be approximated as
\begin{equation}\label{alpha}
Re^{-1}=\alpha (\frac{H}{r})^{2} \leq \frac{\pi a^2}{(G
M_{\ast})^2(1+\qx)^{2/3}n^{4/3}{\sum}_{ml}\frac{m^3}{m-1}|\phiml|^2},
\end{equation}
where the radius of the inner Lindblad resonance is
$r=((m-1)/l)^{2/3}(1+\qx)^{-1/3}a$, $\alpha$ is the
Shakura-Sunyaev viscosity parameter, $a$ is the semmimajor axis of
the binary orbit, and $\qx\equiv \Mx/M_{\ast}$. The Reynolds
number $Re$ is determined by the thermal structure of the disk.

In our calculation, an isothermal disc with $\cs/\vk(R_{\ast})\sim
4.1\times 10^{-2}(\Td/\Teff)^{1/2}$ is assumed as in
\citet{oka01}, where $\cs$ is the sound speed of the disk, $\vk$
is the Keplerian velocity, and the disk temperature $\Td$ is about
$1/2$ of the effective temperature $\Teff$ of the Be star. For a
Keplerian disk, the scale height $H$ of the disk is then
\begin{equation}
\frac{H}{r}=\frac{\cs}{\vk(R_{\ast})}(\frac{r}{R_{\ast}})^{1/2}.
\end{equation}
We adopt $R_{\ast}/R_{\sun}\approx (M_{\ast}/M_{\sun})^{0.8}$ to
get the corresponding radius of the donor star.

The key function in the criterion (\ref{alpha}) is the potential
component $\phiml$, which can be expressed as
\begin{equation}\label{phiml}
\phiml=-\frac{G \Mx}{a}\frac{1}{\pi}{\int_{0}}^{\pi}df
\frac{(1-e^2)^{1/2}}{1+e\cos
(f)}\cos(mf-n(m-1)M(f))b_{1/2}^m(r/r_2(f)),
\end{equation}
where
\begin{equation}
b_{1/2}^m(x)\equiv \frac{2}{\pi}\int_{0}^{\pi}\frac{\cos m\varphi
d\varphi}{\sqrt{1-2x\cos \varphi+x^2}}
\end{equation}
is the Laplace coefficient with argument $x=r/r_2$, and
$r_2=a(1-e^2)/(1+e\cos (f))$ is the distance of the compact star
from the donor star. To make it convenient to solve the
integration in $\phiml$, we have used the following relations
among the mean anomaly $M$, the true anomaly $f$ and the eccentric
anomaly $E$ in an orbital ellipse,
\begin{equation}\label{dMdf}
dM/df=(1-e^2)^{3/2}/(1+e\cos(f))^2 \,
\end{equation}
\begin{equation}\label{ME}
M=E-e\sin E \,
\end{equation}
\begin{equation}\label{Ef}
\sin E=\frac{r_2\sin f}{a\sqrt{1-e^2}}=\frac{\sqrt{1-e^2}}{1+e\cos
f }\sin f .
\end{equation}
From Eqs. (\ref{ME}) and (\ref{Ef}), we get
\begin{equation}\label{Mf}
M(f)=\cases {\arcsin (z(f))-e \cdot z(f),\   \ {\textrm{if}}\  \
0\leq f<\arccos(-e); \cr {(\pi-\arcsin (z(f)))-e \cdot z(f),\   \
{\textrm{if}}\
 \arccos(-e)\leq f \leq \pi}}
\end{equation}
where $z(f)=(1-e^2)^{1/2}\sin f/(1+e\cos f )$. Inserting $M(f)$ to
equation (\ref{phiml}), the numerical integration can be done
directly.

In the calculation of the truncation criterion (\ref{alpha}),
$l=n(m-1)$ for the inner Lindblad resonance, the parameter of the
summation is only $m$ then. Because the high order components
contribute little, we sum the inner Lindblad resonance torque from
$m=2$ to the value at which the component is three orders smaller
than that of $m=2$. The truncation radius deduced from the inner
Lindblad radius is then
\begin{equation}\label{rtrunc}
\rtrunc=n^{-2/3}(1+\qx)^{-1/3}a =
n^{-2/3}(GM_{\ast}/(2\pi/\Porb)^2)^{1/3}.
\end{equation}

The timescale to open the gap between $\rtrunc$ and the inner
Lagrangian point $\dlo$ with $\Delta r=\dlo-\rtrunc$, is about
$t_{\rm open}\approx Re_{\rm {crit}}({\Delta
r}/{\rtrunc})^2\Porb/2\pi$ as in \citet{art94}, where $Re_{\rm
crit}=\alpha_{\rm crit}^{-1} ({H}/{r})^{-2}$ denotes the Reynolds
number for which the gap is opened from $\rtrunc$. Since the
outflow in the disc is subsonic \citep{oka01}, the timescale
$\tau_{\rm drift}\sim \Delta r/\velr$ of a particle drifting from
$\rtrunc$ to the Roche lobe will be longer than the truncation
timescale. Thus, the efficient truncation is defined as
\citep{oka01}
\begin{equation}\label{gamma}
\gamma \equiv (\tau_{\rm drift}/\Porb)_{\rm min}\sim \Delta
r/(0.1\cs\Porb)>1 \,
\end{equation}
where $\velr_{\rm max}\sim 0.1 \cs$ and $\dlo=(0.500-0.227\lg
\qx)a(1-e)$ - the distance of the inner Lagrangian point point
from the center of the donor \citep{fra02} at periastron have been
used instead of the Roche radius, since it is the flat disk rather
than the star itself expanding to the Roche lobe.

\section{Result}

To study the influence of mass ratio, orbital period, eccentricity
and viscosity coefficient on the efficiency of disc truncation, we
have calculated the cases with $\Mx=1.4, 10\, M_{\sun}$,
$M_{\ast}=6,15,20\, M_{\sun}$, $\Porb=5,30,100,250\, {\rm d}$,
$\alpha=0.03,0.1,0.3$ and $e=0.01,0.05,0.1,0.3,0.5,0.7$. The
masses of the Be star and the orbital periods are chosen according
to Fig. 2. of \citet{pod03} and the observations of HMXBs. It is
shown in their figure that most BH X-ray binaries are born in
orbits with $\Porb<30 \, {\rm d}$ and $M_{\ast}<15 \, M_{\sun}$.
On the other hand, most observed Be/NS binaries have the donor
stars with masses of about $10-20 \, M_{\sun}$, and orbital
periods lying in the range of $16 \, {\rm d}<\Porb<300 \, {\rm
d}$. The viscosity coefficient and the eccentricity are adopted
for convenient comparison with the result in Table 2 of
\citet{oka01}.

The calculated values of the truncation efficiency criterion
$(\tau_{\rm drift}/\Porb)_{\rm min}$ are shown against the
eccentricity in Figs. 1, 2 and 3 with $M_{\ast}=6,15,20\,
M_{\sun}$ respectively. The data of different $\Porb$ are grouped.
The general features of the figures
can be summarized as follows:\\
(1) the smaller viscosity parameter of the disk, the more
efficient truncation (it can be understood from equation
(\ref{trunc}) that, given the resonance torques by the compact
star, discs of weak viscosity are truncated at small radius);\\
(2) the narrower systems, the more efficient truncation, since $\gamma\propto a/\Porb\propto \Porb^{-1/3}$;\\
(3) for systems with $e>0.1$, smaller $e$
leads to more efficient truncation, because of the corresponding larger distance to the inner lagrangian point at periastron;  \\
(4) the larger mass of the compact star have a little less
effective truncations on the disk of Be star (since $\dlo$ of Be
star is smaller for more massive companion stars, whereas
$\rtrunc$ has no relation to $\Mx$ as shown in equation
(\ref{rtrunc})); discs of more massive Be stars would be truncated
more effectively, especially when $e<0.4$.\\

Features (1) and (3) have been found in \citet{oka01}, and the
viscous effect has been studied in detail by 3D SPH simulations in
\citet{oka02}. However, feature (2) was neglected. It is shown in
our figures that the influence of the orbital period on the disk
truncation
 is the most promising one among the parameters. Although feature (4)
 shows that black holes would not have more
effective truncation on the decretion disk of the Be stars than
neutron stars, with the help of feature (2), we know that if Be/BH
binaries are mostly born in narrow systems, effective truncation
would still happen (except for very eccentric systems), so that
the accretion would be mainly from the polar wind of the donor
star.

\section{Discussion and Conclusion}

We have extended the application of Be disk truncation model
proposed by \citet{neg01} and \citet{oka01} to general Be binary
systems, to study the dependance of truncation efficiency on some
basic binary parameters, such as the stellar mass, viscosity
coefficient, orbital period and eccentricity. The results not only
confirm the trend that low viscosity and small eccentricity would
lead to effective Be disk truncation \citep{oka01}, but also show
that most effective truncation would occur in relatively narrow
systems.

The above results can help us to understand the observations of Be
binaries. In the context of Be/NS binaries, the observed transient
features of the wide ($\Porb\sim 17-263\ \rm {d}$) eccentric
($e\sim 0.1-0.9$) systems can be well understood by the disk
truncation effect. Furthermore, the results also presents a
possible explanation for the absence of Be/BH binaries if they are
formed in relatively narrow orbits without high eccentricity, as
suggested by \citet{pod03}. The relatively narrow orbit of
$\Porb<30 \, {\rm d}$ is supported by the result of binary
population synthesis simulation work of \citet{pod03} with the
assumption of circular orbit. The knowledge of the eccentricity
could be traced back to the formation of the BHs in the massive
binaries. The BHs were born from the collapse of massive stars in
two different ways (see \citet{fry99} and reference therein ). (i)
The massive stars directly collapse into black holes. (ii) After
supernova explosion, a significant part of the stellar matter
falls back onto the proto-neutron star forming a black hole. If
there were no kick in the latter case, the eccentricity induced by
the supernova can be estimated by $e=\Delta M/\Mf$ \citep{bha91},
where $\Delta M$ is the total mass loss from the binary in the
supernova explosion, and $\Mf$ is the final mass of the binary.
Assuming the mass of the secondary star change little during the
process, one can expect less $\Delta M$ and larger $\Mf$ in black
hole binaries, resulting in smaller eccentricity compared to
neutron star binaries. The spatial distribution and the kinematics
of most of the black-hole binaries appear to be consistent with
the assumption that small asymmetric kicks are imparted to the
black holes \citep{whi96}, with at least one exception GRO
J1655-40, the first black hole system which has evidence of
runaway space velocity of $112\pm 18 \, \rm {km\, s^{-1}}$ in an
orbit of $e=0.34\pm 0.05$ \citep{mir02}.

Because of very effective tidal truncation on Be disks in the
relatively narrow and low eccentric systems, the flow of the
matter towards the BH will be effectively blocked during almost
the whole orbital cycle. In such episode, the accretion could be
mainly from the polar wind of the donor star. As suggested by
\citet{wat89}, the polar wind of the Be star probably resembles
that in OB stars. Such low-density high-velocity wind can hardly
form accretion disks around the BHs, and the accretion would
follow the classical Bondi-Hoyle-Littleton (BHL) approximation.
For a typical main-sequence $15 \, M_{\sun}$ main-sequence star in
the system of $\Porb=10 \, {\rm d}$, if the mass loss rate
$\Mwdot=10^{-9} \, M_{\sun}\ {\rm yr}^{-1}$ and the terminal wind
velocity $v_{\infty}=2000 \, {\rm km \, s}^{-1}$ exist, the
accreting compact star would have luminosity of about $\Lx\sim
10^{33} \, \rm{erg \, s^{-1}}$. For the Be/BH binaries, the
luminosity would be still less, since the polar wind just takes a
small fraction of the mass loss from the donor star. The wind
plasma accumulates in the outer rings of the decretion disc till
one-armed oscillaiton instability - probably responsible for the
V/R variability seen in the Be stars - occurs, which causes the
almost total disruption of the Be disk. The resulted large
mass-infall onto the black hole would lead to very luminous
outbursts. However, no aspect of the truncation model implies that
the large mass transfer must occur \citep{oka02}. If the collapsed
disk due to the dynamical instability falls back on to the more
massive Be star, such narrow and low-eccentricity transients would
be almost out of detection even for the most sensitive
instruments, so it should be careful to deal with the Be stars
unrelated to the X-ray sources.

For systems in which effective truncation occurs, e.g. XTE
J$1543-568$ and 2S $1553-542$, the burst activities are very rare
(see \citet{oka02} and reference therein). During the long
quiescent stage, Be stars could be identified from the Balmer (and
sometimes other) line emission and the associated strong infrared
excess. In the HMXB catalogue of \citet{liu00}, there are probably
24 Be/X-ray systems where the nature of the compact star and the
orbital period are undetermined. They are signed as Be/X-ray
binaries because of their highly variable X-ray characteristics
analogous to those of the well studied Be/NS transients, or the
optical identification of the donor stars. Most of them have been
observed once during the burst states, and then disappeared from
X-ray detection because of very low luminosity. Among these
undetermined Be/X-ray systems, XTE J$1739-302$ and AX
J$0052.9-7158$ (SMC 32) can be excluded now since the donor star
of the former has been identified as an O supergiant \citep{smi03}
and 167.8 seconds pulsations have been found in the latter
\citep{yok01}. To look for Be/BH binaries in the left systems,
optical observations in the quiescent stage become important, from
which we can confirm the nature of the donor star and get
information of the orbital period and the velocity curve of the
donor. The latter two are useful in estimating the dynamical mass
of the compact star - the common way to determine whether the
compact star is a black hole, though the measurements would be
very difficult in Be/X-ray binaries, because of not only the
relatively low velocity of the donor but also the variability of
the circumstellar envelope of Be stars \citep{rag99}. The accurate
locations of some of the undetermined Be/X-ray binaries could be
derived with Chandra, which will enable following up observations
in the infrared/optical wavelength. Till now, we have found very
limited data of these faint Be/X-ray binaries, and there is at
least one system GRS $1736-297$ went undetected even in the
Chandra observations \citep{wil03}.

Even if the X-ray outbursts due to the large mass inflow from the
erupted disks are observed, it is still difficult to distinguish
between black holes and neutron stars, unless pulsations have been
found in the latter. Although there is growing evidence that black
hole X-ray binaries display radio emission when they are in the
low/hard X-ray state \citep{reig03}, this is not a decisive
criterion. LS I$+61{\degr}303$ is an example, whether the compact
star is a neutron star \citep{zam95} or a magnetized black hole
\citep{pun99} is still in debate.

Be stars can be distinguished from normal O/B stars by the special
dense disk wind structures when they are on the main sequence.
Once they evolve to fill the Roche lobe, they can hardly be
distinguished. Due to the high mass transfer rate from the Roche
lobe overflow, black holes in HMXBs would be extremely luminous,
even well above the Eddington limit $\LEdd$ for a stellar-mass
black hole, resulted from either the beaming effect \citep{kin01}
and/or being genuinely super-Eddington \citep{beg02}, and thus
become ones of the ultra luminous X-ray sources (ULXs). This can
also be seen in fig. 6. of \citet{pod03}.

Finally, we conclude that by investigating the disk truncation
efficiency in Be/compact star binary systems, we propose a
possible explanation for the absence of the observation of Be/BH
binaries. We show that if most of the Be/BH binaries are born in
relatively short low eccentric orbit, the interaction of the
binary stars will lead to very effective truncation on the
decretion disk around the Be star, so that most of the Be/BH
systems would appear as very low luminous X-ray sources.

\acknowledgments {We thank Atsuo Okazaki for discussions on the
calculation of the Be disk truncation model, and Dr. X.-L. Luo and
Dr. Zhi Xu for their help in preparing the figures. We acknowledge
the support from NSFC grant 1007003 and NKBRSF-G19990754 grant.}

\clearpage

\begin{figure}
\plottwo{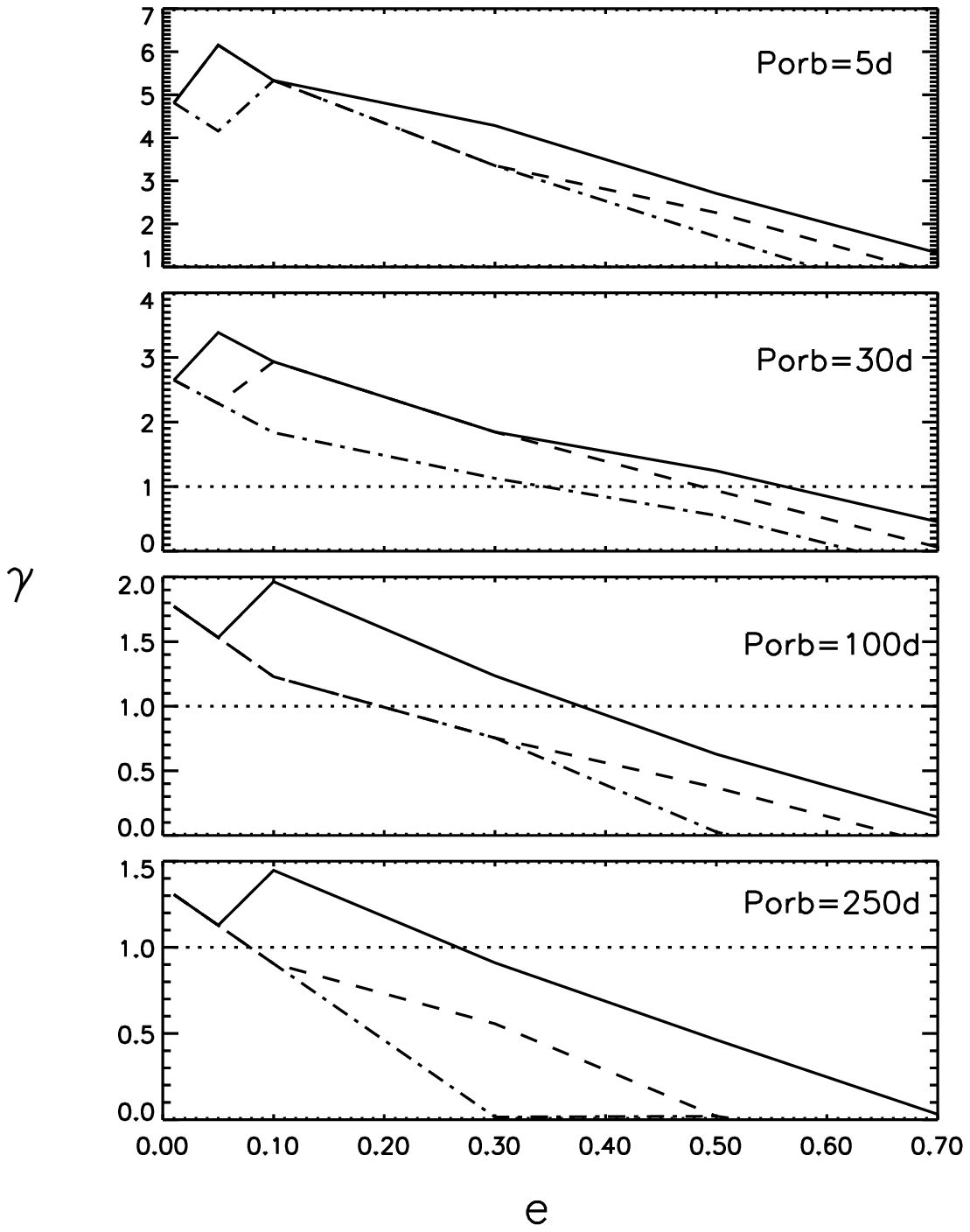}{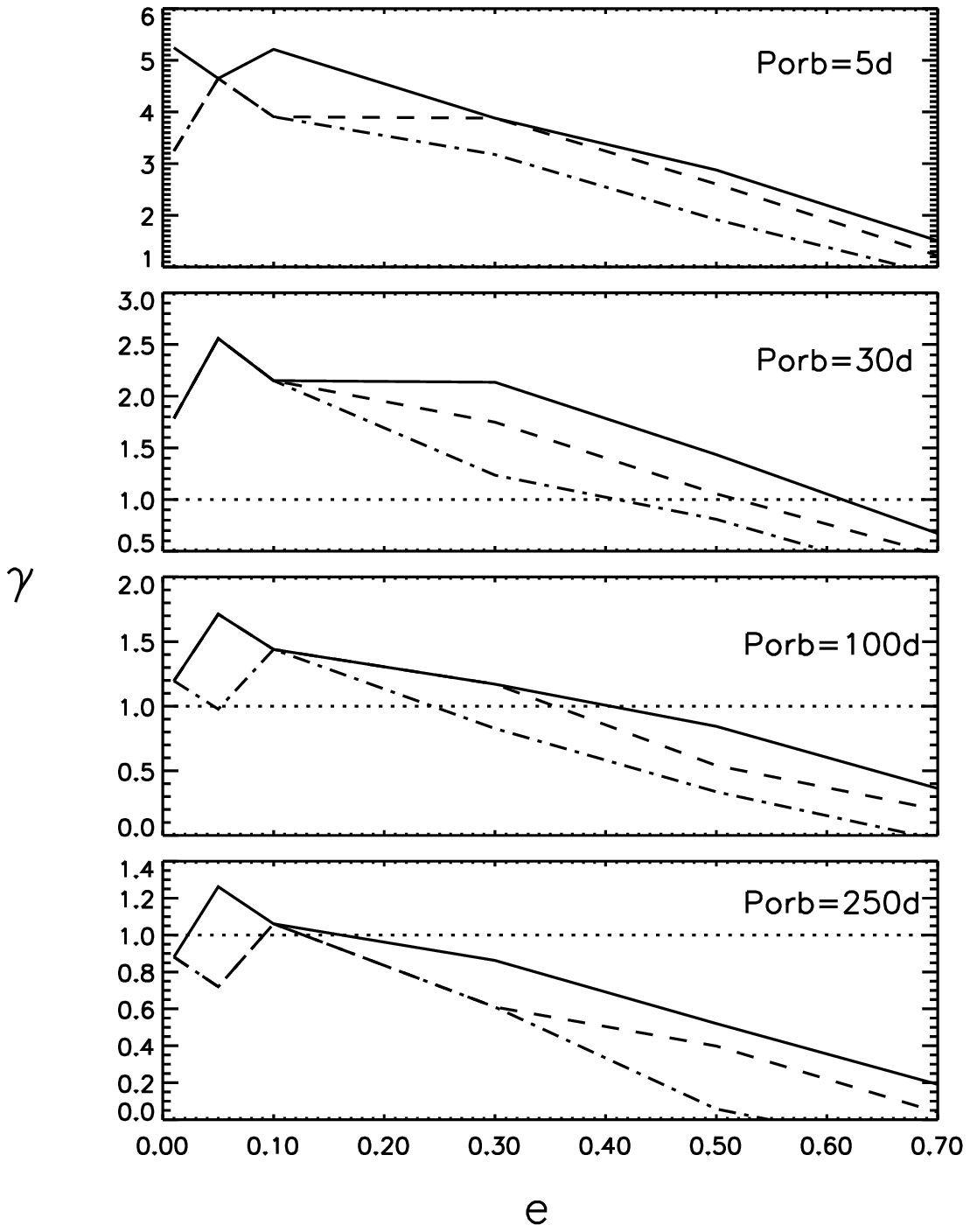} \caption{Eccentricity dependence of the
truncation efficiency $\gamma\equiv(\tau_{\rm drift}/\Porb)_{\rm
min}$ for Be/neutron star binaries (left) and Be/black-hole
binaries (right) with Be star mass $M_{\ast}=6\, M_{\sun}$. The
full lines, dashed lines and dash-dotted lines are obtained with
the disk viscosity parameter $\alpha=0.03,0.1,0.3$, respectively.
Dotted lines of $\gamma=1$ are drawn for comparison. We delete the
point of $\gamma(e)<0$, since the negative values represent the
ineffective truncation with $\rtrunc>\rlo$ and thus have no
physical meaning.}\label{fig1}
\end{figure}
\clearpage

\begin{figure}
\plottwo{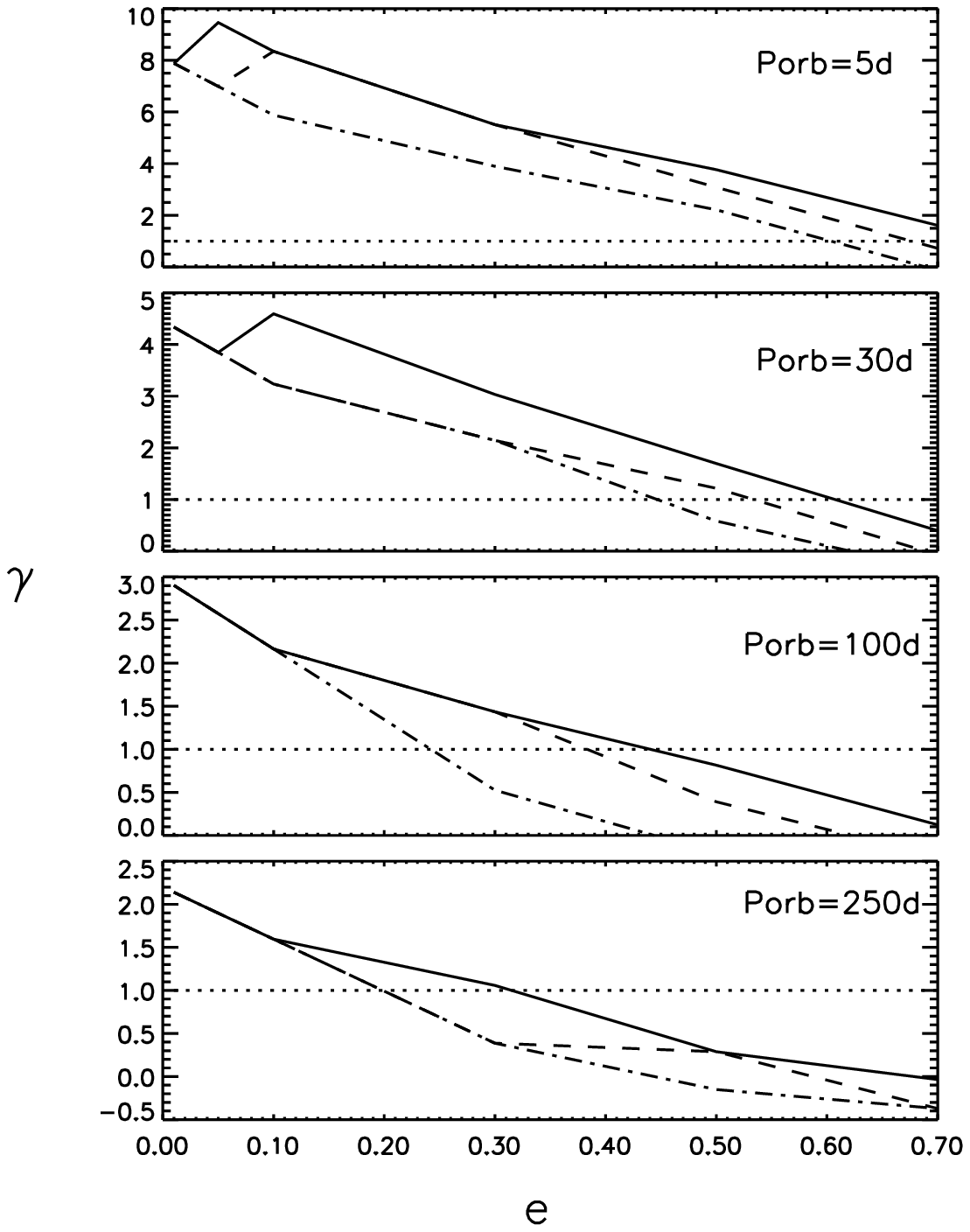}{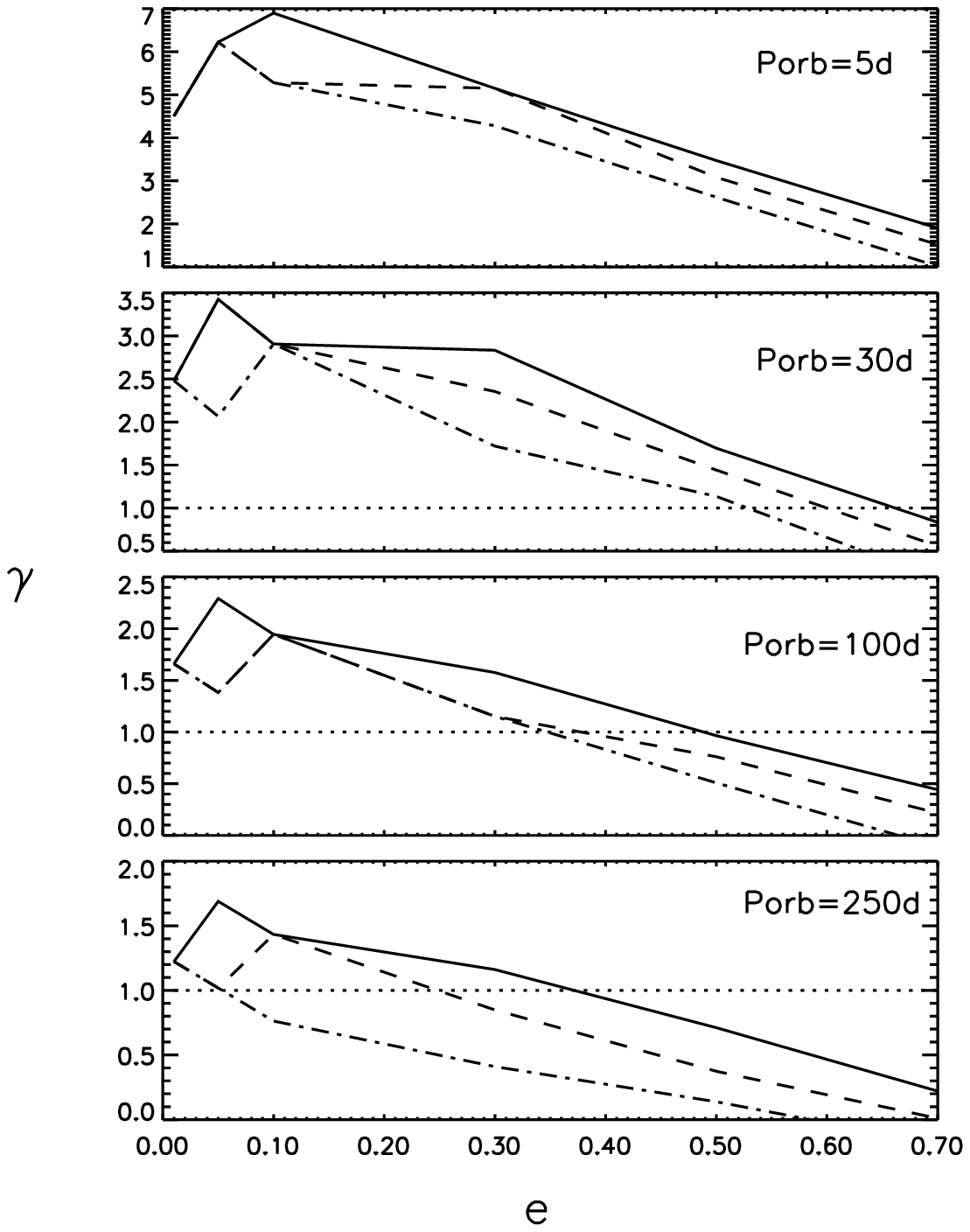} \caption{Same as in Fig.1, but with
$M_{\ast}=15\, M_{\sun}$.} \label{fig2}
\end{figure}

\clearpage

\begin{figure}
\plottwo{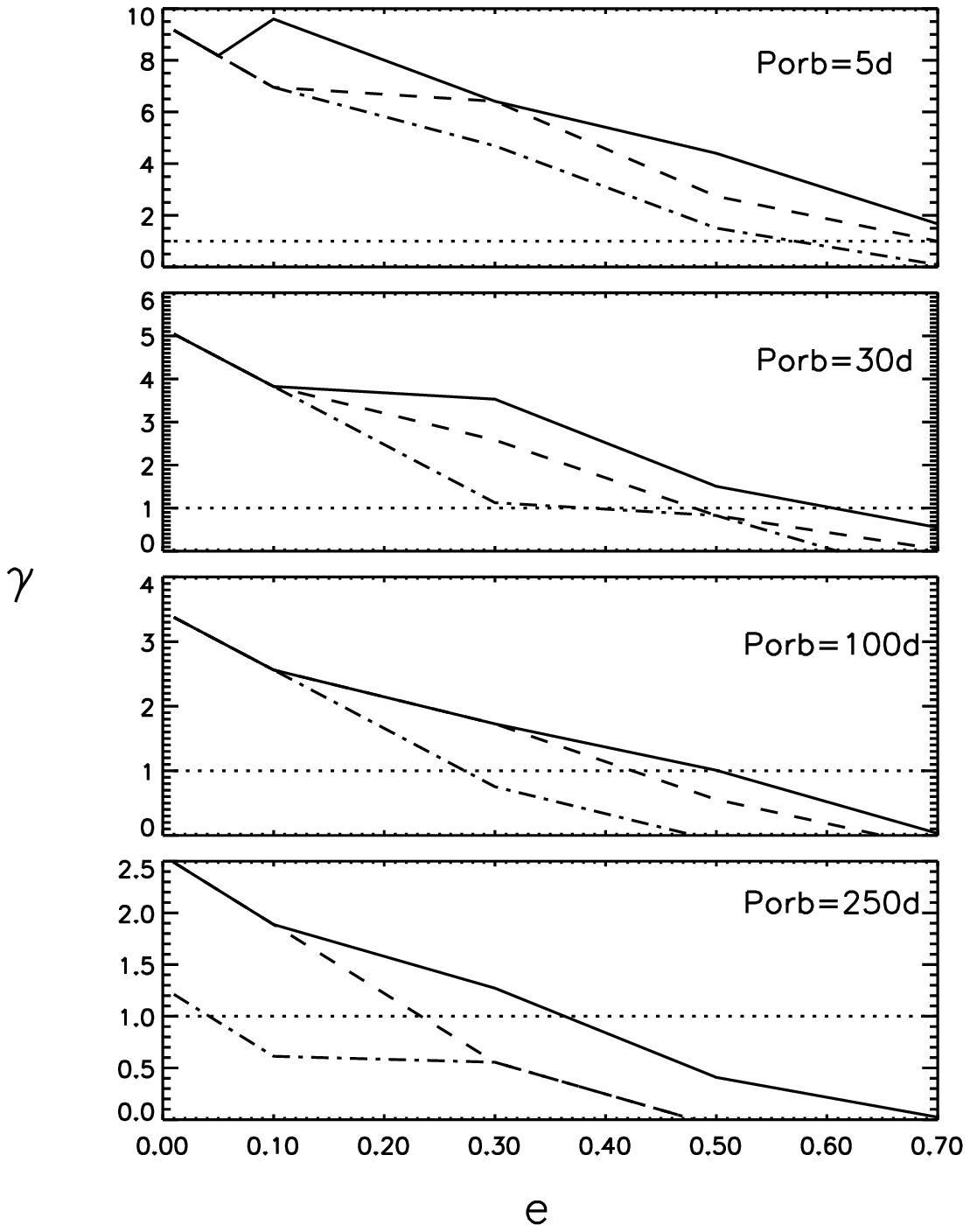}{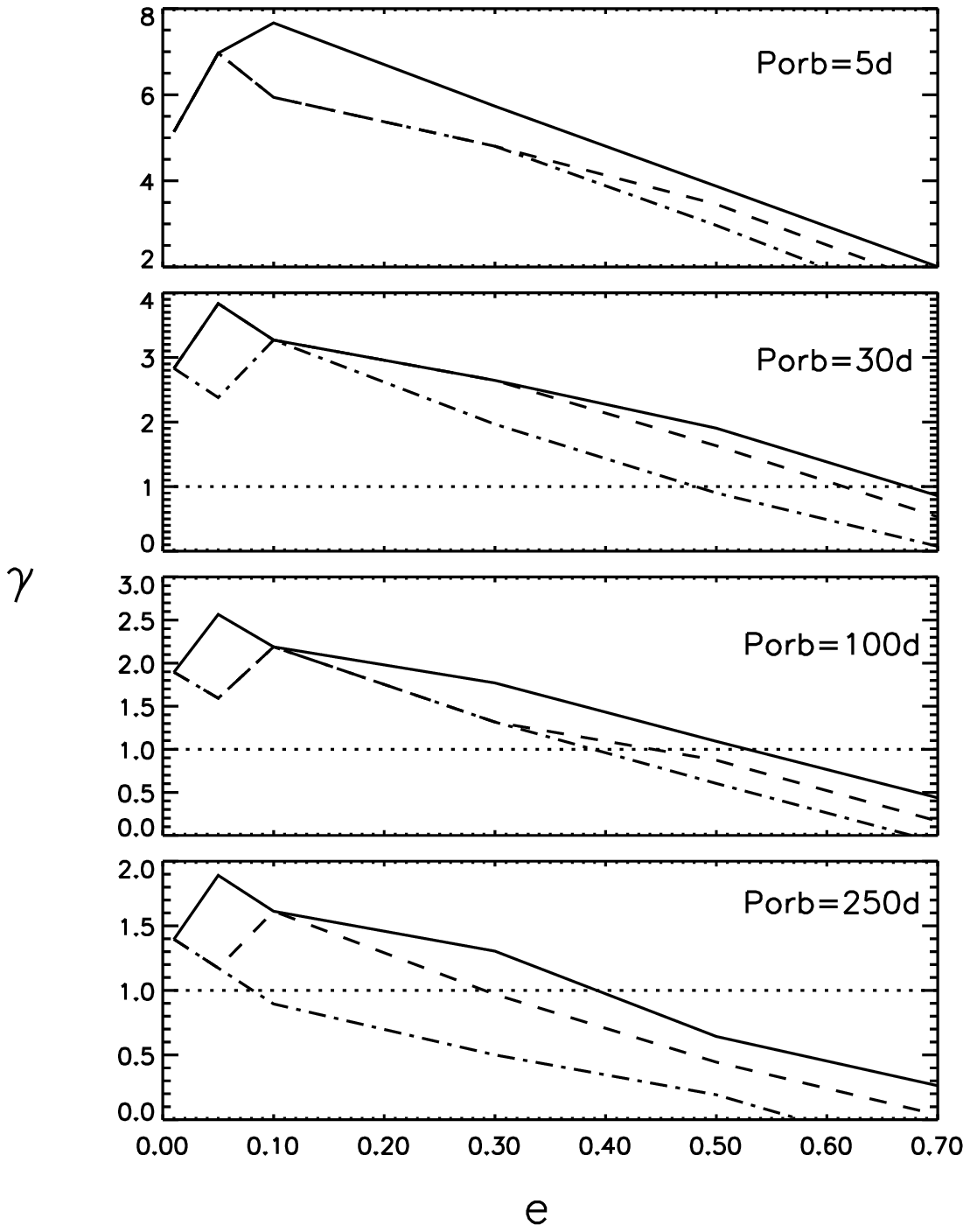} \caption{Same as in Fig.1, but with
$M_{\ast}=20\, M_{\sun}$.}
 \label{fig3}
\end{figure}
\clearpage


\begin{thebibliography}{}
\bibitem[Artymowicz \& Lubow(1994)]{art94} Artymowicz, P., \&
Lubow, S.H. 1994, \apj, 421, 651
\bibitem[Begelman(2002)]{beg02} Begelman, M.C. 2002, \apjl, 568,
L97
\bibitem[Bhattacharya and van den Heuvel(1991)]{bha91}
Bhattacharya, D., \& van den Heuvel, E.P.J. 1991, \physrep 203, 1
\bibitem[Coe(2000)]{coe00} Coe, M.J. 2000, The Be Phenomenon in Early-type Stars, IAU Colloquium 175, ASP Conference Proceedings, Vol. 214, eds. Myron A. Smith and Huib F. Henrichs, (Astronomical
                     Society of the Pacific), 656
\bibitem[Frank, King \& Raine(2002)]{fra02} Frank, J., King, A.R., \& Raines, D.J. 2002,
Accretion Power in Astrophysics (3rd ed)
\bibitem[Fryer(1999)]{fry99} Fryer, C.J. 1999,
 \apj, 522, 413
\bibitem[Fryer \& Kakogera (2001)]{fry01} Fryer, C.J., \& Kalogera, V. 2001, \apj, 554,
548
\bibitem[Greiner et al.(2001)]{gre01} Greiner, J., Cuby, J.G.,
McCaughrean, M.J., Castro-Tirado, A.J., \& Mennickent, R.E. 2001,
\aap, 373, L37
\bibitem[Haigh, Coe \& Fabregat(2003)]{hai03} Haigh, N.J., Coe, M.J., \& Fabregat, J.
 2003, submitted to \mnras, astro-ph/0305194
\bibitem[Kaper et al.(1995)]{kap95} Kaper,L., Lamers, H.J.G.L.M., Ruymaekers, E., et al., 1995, \aap 300, 446
\bibitem[King et al.(2001)]{kin01} King, A.R., Davies, M.B., Ward, M.J., Fabbiano, G., \& Elvis, M. 2001 \apjl, 552,
L109
\bibitem[Lee, Saio \& Osaki(1991)]{lee91} Lee, U., Saio, H., \& Osaki, Y. 1991, \mnras, 250,
432
\bibitem[Liu, Paradijs \& van den Heuvel(2000)]{liu00} Liu, Q.Z., van Paradijs, J., \& van den Heuvel, E.P.J. 2000, \aap, 147, 25
\bibitem[McClintock \& Remillard(2003)]{mcc03} McClintock, J.E. \&
Remillard, R.A. 2003, to appear in Compact Stellar X-ray Sources,
eds. W.H.G. Lewin and M. van der Klis., astro-ph/0306213
\bibitem[Mirabel et al. (2002)]{mir02} Mirabel, I.F., Mignani, R., Rodrigues, I., Combi, J.A., Rodriguez, L.F., \& Guglielmetti, F. 2002, \aap, 393,
595
\bibitem[Negueruela \& Okazaki(2001)]{neg01} Negueruela, I., \&
Okazaki, A.T. 2001, \aap, 369, 108
\bibitem[Negueruela \& Reig(2001)]{neg4u} Negueruela, I. \& Reig, P. 2001,
\aap, 371, 1056
\bibitem[Okazaki \& Negueruela(2001)]{oka01} Okazaki, A.T., \& Negueruela, I.
 2001, \aap, 377, 1610
\bibitem[Okazaki et al.(2002)]{oka02} Okazaki, A.T., Bate, M.R., Ogilvie, G.I., \& Pringle,
J.E. 2002, \mnras, 337, 9670
\bibitem[Podsiadlowski et al.(2003)]{pod03} Podsiadlowski, Ph., Rappaport, S. \& Han, Z. 2003,
\mnras, 341, 385
\bibitem[Porter (1999)]{por99} Porter, J.M., 1999, \aap, 341, 560
\bibitem[Punsly(1999)]{pun99} Punsly, B. 1999, \apj, 519, 336
\bibitem[Raguzova and Lipunov(1999)]{rag99} Raguzova, N.V., \& Lipunov, V.M. 1999, \aap, 349, 505
\bibitem[Raguzova(2001)]{rag01} Raguzova, N.V. 2001, \apss, 281, 641
\bibitem[Robinson, Ivans \& Welsh (2002)]{rob02} Robinson, E.L., Ivans, I.I., \& Welsh, W.F. 2002,
\apj, 565, 1169
\bibitem[Reig, Kylafis \& Giannios(2003)]{reig03} Reig, P., Kylafis, N.,
\& Giannios, D. 2003, \aap, 403, 15
\bibitem[Smith et al.(2003)]{smi03} Smith, D.M., Heindl, W.A., Harrison, T.E., Swank, J.H., \& Markwardt, C.B. 2003, HEAD, 35,
1733
\bibitem[Waters \& van Kerkwijk(1989)]{wat89} Waters,
L.B.F.M. and van Kerkwijk, M.H. 1989, \aap, 223, 196
\bibitem[White \& van Paradijs(1996)]{whi96} White, N.E. \& van
Paradijs, J. 1996, \apjl, 473, L25
\bibitem[Wilson et al.(2003)]{wil03} Wilson, C.A., Patel, S.K.,
Kouveliotou, C., Jonker, P.G., van der Klis, M., Lewin, W.H.G.,
Belloni, T., \& Mendez, M. 2003, accepted by \apj,
astro-ph/0307045
\bibitem[Yokogawa et al.(2001)]{yok01} Yokogawa, J., Torii, K., Kohmura, T., \& Koyama, K. 2001, \pasj, 53,
9
\bibitem[Zamanov(1995)]{zam95} Zamanov, R. 1995, \mnras, 272, 308
\bibitem[Zamanov et al. (2001)]{zam01} Zamanov, R.K., Reig, P., Marti, J., Coe, M.J., Fabregat,J., Tomov, N.A.,
\& Valchev, T. \aap, 367, 884
\bibitem[Ziolkowski(2002)]{zio02} Ziolkowski, J. 2002, MmSAI, 73, 1038 (astro-ph/0208455)
\end{thebibliography}
\end{document}